\title{Direct S-matrix calculation for diffractive structures and metasurfaces}

\author{Alexey A. Shcherbakov$^{1*}$, Yury V. Stebunov$^{1,2}$, Denis F. Baidin$^1$\\ Thomas K\"{a}mpfe$^3$, and Yves Jourlin$^3$ \\\\ $^{1}$Moscow Institute of Physics and Technology, Dolgoprudniy, Russia \\$^2$GrapheneTek, Skolkovo Innovation Center, Russia \\ $^3$University of Lyon, Laboratory Hubert Curien, Saint-Etienne, France \\$^*$shcherbakov.aa@mipt.ru}

\documentclass[12pt]{article}

\usepackage{graphicx}
\usepackage{amsmath,amssymb,bm}
\usepackage[a4paper,total={6.8in,8in}]{geometry}

\DeclareMathOperator{\sinc}{sinc}
\DeclareMathOperator{\sign}{sign}

\begin{document}
\maketitle

\begin{abstract}
The paper presents a derivation of analytical components of S-matrices for arbitrary planar diffractive structures and metasurfaces in the Fourier domain. Attained general formulas for S-matrix components can be applied within both formulations in the Cartesian and curvilinear metric. A numerical method based on these results can benefit from all previous improvements of the Fourier domain methods. In addition, we provide expressions for S-matrix calculation in case of periodically corrugated layers of 2D materials, which are valid for arbitrary corrugation depth-to-period ratios. As an example the derived equations are used to simulate resonant grating excitation of graphene plasmons and an impact of silica interlayer on corresponding reflection curves.
\end{abstract}

\section{Introduction}

Periodic optical structures ranging from conventional one-dimensional diffraction gratings to metasurfaces attract a lot of attention due to wide optimization possibilities in design of optical response functions. The wavelength scale nature of patterns of the most complex diffractive structures and metasurfaces requires an effort to rigorously solve Maxwell's equations. Concerning various methods capable to suit this task the Fourier space methods \cite{Popov2012} are among the most popular and important ones. They are widely used owing to their relatively simple formulation, versatility in structures that can be modelled, and an output in form of S-matrices. The latter property features means that these methods can be directly used in simulation of measurable quantitites. Not only are S-matrices physically important entities, but also they bring stability in numerical calculations \cite{Inkson1988,Sambles1995,Li1996-1,Gippius2009}.

Fourier methods, however, conventionally operate with T-matrices, and to ensure stability of calculations an additional effort is needed, e.g., proposed in \cite{Moharam1995,Sambles1995}. In this paper we develop a way to overcome these complications and demonstrate how S-matrix components of a slice of an arbitrary diffractive structure can be derived analytically in the Fourier space. Our derivation is based on the integral equation solution of the Maxwell's equations, albeit the distribution formalism given in \cite{Sipe1987} can be used to get the same result.

In addition to derivation of S-matrices of bulk material gratings it is shown that expressions for corrugated layers of 2D materials can be attained in a similar way, which provides the possibility to efficiently simulate complex metasurfaces. This widens the applicability of the previous models of \cite{Bludov2013,Nikitin2013} where an electrodynamic response of a graphene sheet placed on top of a corrugated substrate was simulated by means of the Rayleigh-type methods. Our direct S-matrix approach does not rely on the Rayleigh hypothesis and hence is free of convergence issues which the Rayleigh-type methods face \cite{Wauer2009,Tishchenko2010}.


\section{Volume Integral Equation}

Consider a planar structure, which can be either a metasurface or a diffractive optical element, in Cartesian coordinates $X_{\alpha}$, $\alpha = 1,2,3$ with unit vectors $\hat{\bm{e}}_{\alpha}$, such that axis $X_3$ is orthogonal to the structure plane. The structure is supposed to be doubly periodic along two non-collinear directions in plane $X_1X_2$. Denote unit vectors in the directions of the periods as $\hat{\bm{p}}_{1,2}$, and let periods be $\Lambda_{1,2}$. Reciprocal lattice vectors then read
\begin{equation}
	{\bf K}_1 = \frac{2\pi}{\Lambda_1}\frac{\hat{\bm{p}}_2\times\hat{\bm{e}}_z}{\hat{\bm{p}}_{1}\cdot(\hat{\bm{p}}_{2}\times\hat{\bm{e}}_z)}, \; {\bf K}_2 = \frac{2\pi}{\Lambda_2}\frac{\hat{\bm{e}}_z\times\hat{\bm{p}}_1}{\hat{\bm{p}}_{1}\cdot(\hat{\bm{p}}_{2}\times\hat{\bm{e}}_z)}.
	\label{eq:lattice-vectors}
\end{equation}

The paper refers to linear phenomena only. Thus, we consider Maxwell's equations for time-harmonic fields and sources with implicit time dependence exponential factor $\exp(-i\omega t)$, which will be omitted further:
\begin{equation}
	\begin{split}
		&\nabla\times{\bf E} = -{\bf M} + i\omega\mu{\bf H}, \\
		&\nabla\times{\bf H} = {\bf J} - i\omega\varepsilon{\bf E}.
	\end{split}
	\label{eq:Maxwell}
\end{equation}
The magnetic source term $\bf M$ is essential for a curvilinear coordinate formulation as will be discussed below. Eqs.~(\ref{eq:Maxwell}) yield Helmholtz equations providing that the dielectric permittivity and the magnetic permeability are constants. We will refer to these quantities as basis ones and denote them as $\varepsilon_b$, $\mu_b$. Helmholtz equations then read
\begin{equation}
	\begin{split}
		&\nabla\times\nabla\times{\bf E} - k_b^2{\bf E} = i\omega\mu_b{\bf J} - \nabla\times{\bf M}, \\
		&\nabla\times\nabla\times{\bf H} - k_b^2{\bf H} = i\omega\varepsilon_b{\bf M} + \nabla\times{\bf J},
	\end{split}
	\label{eq:Helmholtz}
\end{equation}
where $k_b = \omega\sqrt{\varepsilon_b\mu_b}$ is the wavenumber of the homogeneous basis space. Well-known solutions of Eqs.~(\ref{eq:Helmholtz}) in form of volume integral equations rely on the free-space tensor electric and mixed Green's functions \cite{Felsen1972}, ${\bf G}^e$ and ${\bf G}^m$ respectively:
\begin{equation}
	\begin{split}
	{\bf E}(\bm{r}) &= {\bf E}^{ext}(\bm{r}) + i\omega\mu_b\int d^3\bm{r}'{\bf G}^e(\bm{r}-\bm{r}'){\bf J}(\bm{r}') - k_b\int d^3\bm{r}'{\bf G}^m(\bm{r}-\bm{r}'){\bf M}(\bm{r}'), \\
	{\bf H}(\bm{r}) &= {\bf H}^{ext}(\bm{r}) + i\omega\varepsilon_b\int d^3\bm{r}'{\bf G}^e(\bm{r}-\bm{r}'){\bf M}(\bm{r}') + k_b\int d^3\bm{r}'{\bf G}^m(\bm{r}-\bm{r}'){\bf J}(\bm{r}').
	\end{split}
	\label{eq:E_volume}
\end{equation}
In these equations we single out ''external'' field amplitudes ${\bf E}^{ext}$, ${\bf H}^{ext}$, which are supposed to be known and to be produced by some sources being outside of a region under consideration. In turn, sources present in the right-hand parts of Eqs.~(\ref{eq:E_volume}) will be further on associated with local medium inhomogeneities.

Aimed at dealing with planar structures we utilize a decomposition of the Green's function in the plane wave basis. Given a plane wave wavevector ${\bm{k}}^{\pm} = (k_1,\;k_2,\;\pm k_3)^T$ with sign $\pm$ distinguishing waves propagating upwards and downwards relative to axis $X_3$, whose components meet the dispersion equation $k_1^2+k_2^2+k_3^2=k_b^2$ supplemented with the condition $\Re k_3+\Im k_3>0$ \cite{Petit1980}, the unit vectors of the TE and the TM polarized waves can be taken as
\begin{equation}
	\hat{\bm{e}}^{e\pm} = \frac{{\bm{k}}^{\pm}\times\hat{\bm{e}}_3}{|{\bm{k}}^{\pm}\times\hat{\bm{e}}_3|}, \;\hat{\bm{e}}^{h\pm} = \frac{({\bm{k}}^{\pm}\times\hat{\bm{e}}_3)\times{\bm{k}}^{\pm}}{|({\bm{k}}^{\pm}\times\hat{\bm{e}}_3)\times{\bm{k}}^{\pm}|},
	\label{eq:TE-TM}
\end{equation}
respectively. With this definition the Green's functions explicitly write
\begin{equation}
	\begin{split}
		G^{e}_{\alpha\beta}(\bm{\rho}-\bm{\rho}',x_3-x_3') &= \frac{i}{4\pi^2}\int d^2\bm{\kappa} \left\{ \frac{i}{k_b^2}\delta(x_3-x_3')\delta_{\alpha 3}\delta_{\beta 3} \right. \\ &+ \left. \left[\hat{e}_{\alpha}^{e\sigma}\hat{e}_{\beta}^{e\sigma} + \hat{e}_{\alpha}^{h\sigma}\hat{e}_{\beta}^{h\sigma}\right] \frac{\exp(ik_3|x_3-x_3'|)}{2k_3} \right\}\exp\left[ i\bm{\kappa}(\bm{\rho}-\bm{\rho}') \right]
	\end{split}
	\label{eq:Green_e}
\end{equation}
\begin{equation}
	\begin{split}
		&G^m_{\alpha\beta}(\bm{\rho}-\bm{\rho}',x_3-x_3') = -\frac{1}{4\pi^2}\int d^2\bm{\kappa} \xi_{\alpha\gamma\beta}k^{\sigma}_{\gamma} \frac{\exp(ik_3|x_3-x_3'|)}{2k_3} \exp\left[ i\bm{\kappa}(\bm{\rho}-\bm{\rho}') \right]
	\end{split}
	\label{eq:Green_m}
\end{equation}
Here $\bm{\rho} = (x_1,\;x_2)^T$, $\bm{\kappa}=(k_1,\;k_2)^T$, $\delta_{\alpha\beta}$ and $\xi_{\alpha\beta\gamma}$ are Kronecker and Levi-Civita symbols respectively, and $\sigma=\sign(x_3-x_3')$. Expressions similar to the first equation for the electric Green's functions can be found, e.g., in \cite{Munk1979,Tomas1995,Peres2017}, and derivation of the second is analogous. For consistency the derivation is briefly reviewed in Appendix.

In order to proceed to analysis of periodic structures, first, let us fix the ''zero harmonic'' wavevector $\bm{k}^{(0)} = \left(\bm{\kappa}^{(0)},\sqrt{k_b^2-(\bm{\kappa}^{(0)})^2}\right)^T$ so that fields and sources are subject to Floquet-Bloch condition
\begin{equation}
	{\bf V}(\bm{\rho}+m_1\Lambda_1\hat{\bm{p}}_1 + m_2\Lambda_2\hat{\bm{p}}_2,z) = {\bf V}(\bm{\rho},z) \exp\left[i\bm{\kappa}^{(0)}(m_1\Lambda_1\hat{\bm{p}}_1 + m_2\Lambda_2\hat{\bm{p}}_2)\right]
	\label{eq:Floquet-Bloch}
\end{equation}
Vector ${\bf V}$ can be substituted by any of ${\bf E}$, ${\bf H}$, ${\bf J}$, ${\bf M}$. Under this condition one can apply the Poisson summation formula in Eqs.~(\ref{eq:E_volume}) with explicit functions Eqs.~(\ref{eq:Green_e}), (\ref{eq:Green_m}) to arrive at equations which depend on the Fourier components of the sources, which read
\begin{equation}
	{\bf S}_m(x_3) = \frac{|{\bf K}_1\times{\bf K}_2|}{4\pi^2} \iint\limits_P d^2\bm{\rho}' {\bf S}(\bm{\rho}',x_3)\exp(-i\bm{\kappa}_m\bm{\rho}')
	\label{eq:source_Fourier}
\end{equation}
where ${\bf S}$ stands either for ${\bf J}$ or ${\bf M}$, the integration is performed over one structure period, and the plane harmonic wavevector $\bm{\kappa}_m = \bm{\kappa}^{(0)}+m_1{\bf K}_1+m_2{\bf K}_2$ depends on the two-dimensional index $m=(m_1,\;m_2)\in\mathbb{Z}^2$. The solution in the periodic domain then explicitly writes
\begin{equation}
	\begin{split}
		&{\bf E}(\bm{\rho},x_3) = {\bf E}^{ext}(\bm{\rho},x_3) + \frac{J_3(\bm{\rho},x_3)}{i\omega\varepsilon_b}\hat{\bm{e}}_3 \\
		& - \omega\mu_b \sum\limits_m \exp(i\bm{\kappa}_m\bm{\rho}) \int\limits_{-\infty}^{\infty} dx_3' \left[ \hat{\bm{e}}_m^{e\sigma}(\hat{\bm{e}}_m^{e\sigma}\cdot{\bf J}_m(x_3')) + \hat{\bm{e}}_m^{h\sigma}(\hat{\bm{e}}_m^{h\sigma}\cdot{\bf J}_m(x_3')) \right] \frac{\exp(ik_{3m}|x_3-x_3'|)}{2k_{3m}} \\
		& + k_b\sum\limits_m \exp(i\bm{\kappa}_m\bm{\rho}) \int\limits_{-\infty}^{\infty} dx_3' \left[ \hat{\bm{e}}_m^{e\sigma}(\hat{\bm{e}}_m^{h\sigma}\cdot{\bf M}_m(x_3')) - \hat{\bm{e}}_m^{h\sigma}(\hat{\bm{e}}_m^{e\sigma}\cdot{\bf M}_m(x_3')) \right] \frac{\exp(ik_{3m}|x_3-x_3'|)}{2k_{3m}},
	\end{split}
	\label{eq:E_volume_periodic}
\end{equation}
and a similar expression holds for the magnetic field. Here $k_{3m}$ are propagation constants of plane harmonics defined by the dispersion equation $\bm{\kappa}_m^2+k_{3m}^2=k_b^2$ and the condition $\Re k_{3m}+\Im k_{3m} > 0$; and $\hat{\bm{e}}_m^{e,h\pm}$ are unit polarization vectors obtained by using wavevectors $\bm{k}^{\pm}_m=(\bm{\kappa}_m,\pm k_{3m})^T$ in Eqs.~(\ref{eq:TE-TM}). Eq.~(\ref{eq:E_volume_periodic}) shows that the electric field is a superposition of plane harmonics and a source ''delta''-term.

Let us introduce the modified field $\tilde{E}_{1,2}=E_{1,2}$, $\tilde{E_3} = E_3-J_3/i\omega\varepsilon_b$, so that the plane wave decomposition of this field
\begin{equation}
	\tilde{\bf E}(\bm{\rho},x_3) = \sum_m \left[ \tilde{a}_m^{e+}(x_3)\hat{\bm{e}}_m^{e+} + \tilde{a}_m^{e-}(x_3)\hat{\bm{e}}_m^{e-} + \tilde{a}_m^{h+}(x_3)\hat{\bm{e}}_m^{h+} + \tilde{a}_m^{h-}(x_3)\hat{\bm{e}}_m^{h-} \right] \exp(i\bm{\kappa}_m\bm{\rho})
	\label{eq:modified-field-amp}
\end{equation}
is valid at any space point. Modified amplitude $\tilde{\bf E}^{ext}$ of the external field is identical to non-modified ${\bf E}^{ext}$ since the sources of this field are supposed to be outside the region of interest, and we intend to evaluate Eq.~(\ref{eq:E_volume_periodic}) outside these sources. We assume then that a decomposition similar to Eq.~(\ref{eq:modified-field-amp}) for the external field amplitudes is known yielding amplitudes $\tilde{a}_m^{ext,e\pm}$ and $\tilde{a}_m^{ext,h\pm}$. Combining Eqs.~(\ref{eq:E_volume_periodic}) and (\ref{eq:modified-field-amp}) we see that the unknown amplitudes are found from integration over the third coordinate
\begin{equation}
	\begin{split}
		&\tilde{a}_m^{e+}(x_3) = \tilde{a}_m^{ext,e+}(x_3) - \int\limits_{-\infty}^{x_3} d\zeta \frac{\exp[ik_{3m}(x_3-\zeta)]}{2k_{3m}} \left[ \omega\mu_b \hat{\bm{e}}_m^{e+}\cdot{\bf J}_m(\zeta) - k_b \hat{\bm{e}}_m^{h+}\cdot{\bf M}_m(\zeta) \right],\\
		&\tilde{a}_m^{h+}(x_3) = \tilde{a}_m^{ext,h+}(x_3) - \int\limits_{-\infty}^{x_3} d\zeta \frac{\exp[ik_{3m}(x_3-\zeta)]}{2k_{3m}} \left[ \omega\mu_b \hat{\bm{e}}_m^{h+}\cdot{\bf J}_m(\zeta) + k_b \hat{\bm{e}}_m^{e+}\cdot{\bf M}_m(\zeta) \right],\\
		&\tilde{a}_m^{e-}(x_3) = \tilde{a}_m^{ext,e-}(x_3) - \int\limits^{\infty}_{x_3} d\zeta \frac{\exp[ik_{3m}(\zeta-x_3)]}{2k_{3m}} \left[ \omega\mu_b \hat{\bm{e}}_m^{e-}\cdot{\bf J}_m(\zeta) - k_b \hat{\bm{e}}_m^{h-}\cdot{\bf M}_m(\zeta) \right],\\
		&\tilde{a}_m^{h-}(x_3) = \tilde{a}_m^{ext,h-}(x_3) - \int\limits^{\infty}_{x_3} d\zeta \frac{\exp[ik_{3m}(\zeta-x_3)]}{2k_{3m}} \left[ \omega\mu_b \hat{\bm{e}}_m^{h-}\cdot{\bf J}_m(\zeta) + k_b \hat{\bm{e}}_m^{e-}\cdot{\bf M}_m(\zeta) \right].
	\end{split}
	\label{eq:amp-src}
\end{equation}
These equations can be used to get a formulation of the Generalized Source Method either in Cartesian \cite{Shcherbakov2012} or in curvilinear \cite{Shcherbakov2017} coordinates by introducing Generalized Sources related to the fields and performing implicit numerical integration. Instead, in the next section Eqs.~(\ref{eq:amp-src}) are used to obtain analytical S-matrix components of a thin grating layer.


\section{S-matrix of a thin grating slice}

On the way towards analytical S-matrix components we associate a region of interest, or source region, for general solution given by Eqs.~(\ref{eq:amp-src}) with a plane layer $x_3^{(1)}\leq x_3\leq x_3^{(2)}$ of thickness $\Delta x_3 = x_3^{(2)}-x_3^{(1)}$. If $\Delta x_3\rightarrow 0$ the integration in Eqs.~(\ref{eq:amp-src}) is reduced to a multiplication of integrands by $\Delta x_3$. Let us denote coordinate of layer center as $x_3^{c}=\left(x_3^{(1)}+x_3^{(2)}\right)/2$. Once incoming wave amplitudes at layer boundaries are known, Eqs.~(\ref{eq:amp-src}) yield the diffracted amplitudes in form
\begin{subequations}
\begin{equation}
	\begin{split}
		\tilde{a}_m^{e\pm}\left(x_3^c \pm \Delta x_3/2\right) & \approx \tilde{a}_m^{ext,e\pm}\left(x_3^c \pm \Delta x_3/2\right) \\
		& - \Delta x_3 \frac{\exp(ik_{3m}\Delta x_3/2)}{2k_{3m}} \left[ \omega\mu_b \hat{\bm{e}}_m^{e\pm}\cdot{\bf J}_m(x_3^c) - k_b \hat{\bm{e}}_m^{h\pm}\cdot{\bf M}_m(x_3^c) \right],\\
	\end{split}
	\label{eq:amp-src-approx_e}
\end{equation}
\begin{equation}
	\begin{split}
		\tilde{a}_m^{h\pm}\left(x_3^c \pm \Delta x_3/2\right) & \approx \tilde{a}_m^{ext,h\pm}\left(x_3^c \pm \Delta x_3/2\right) \\
		& - \Delta x_3 \frac{\exp(ik_{3m}\Delta x_3/2)}{2k_{3m}} \left[ \omega\mu_b \hat{\bm{e}}_m^{h\pm}\cdot{\bf J}_m(x_3^c) + k_b \hat{\bm{e}}_m^{e\pm}\cdot{\bf M}_m(x_3^c) \right],\\
	\end{split}
	\label{eq:amp-src-approx_h}
\end{equation}
\label{eq:amp-src-approx}
\end{subequations}
Integration in the first pair of Eqs.~(\ref{eq:amp-src}) is performed up to $x_3=x_3^c + \Delta x_3/2$, and in the second pair -- from $x_3=x_3^c - \Delta x_3/2$. Sources that produce fields with amplitudes $\tilde{a}_m^{ext}$ can be located anywhere outside the layer.

To get closed form equations $\bf J$ and $\bf M$ should be related to the fields. In the simplest case one may take ${\bf J} = -i\omega(\varepsilon-\varepsilon_b){\bf E}$, and ${\bf M}=0$, as is done within the Generalized Source Method for gratings whose permittivity is represented by smooth spatial functions \cite{Shcherbakov2012} or in other Volume Integral Equation methods (e.g., \cite{Martin1998}). When a formulation includes a correct treatment of the boundary conditions to a corrugation interface in the Fourier domain \cite{Li1996} or the Generalized Metric Sources, which appear in a curvilinear formulation of the problem \cite{Shcherbakov2013}, the dependence of the sources from the field can be more complicated. Generally, such dependence writes as
\begin{equation}
	\begin{split}
		&{\bf J} = -i\omega\varepsilon_b\Omega_E{\bf E} \\
		&{\bf M} = -i\omega\mu_b\Omega_H{\bf H} \\
	\end{split}
	\label{eq:source-field}
\end{equation}
with $3\times3$ matrices $\Omega_{E,H}$, whose explicit form is supposed to be known, but is not needed for the derivation in this section.

In order to operate with plane harmonic amplitudes only we introduce matrices composed of column vectors given by Eq.~(\ref{eq:TE-TM}):
\begin{equation}
	\begin{split}
		&\mathcal{V} = \left\{\hat{\bm{e}}^{e+},\;\hat{\bm{e}}^{h+},\;\hat{\bm{e}}^{e-},\;\hat{\bm{e}}^{h-}\right\}, \\
		&\mathcal{W} = \left\{\hat{\bm{e}}^{h+},\;-\hat{\bm{e}}^{e+},\;\hat{\bm{e}}^{h-},\;-\hat{\bm{e}}^{e-}\right\}.
	\end{split}
	\label{eq:matrix-V}
\end{equation}
Additionally, let us denote amplitude vectors of upward and downward propagating harmonics as $\tilde{\bm{a}}_m^{\pm} = (\tilde{a}_m^{e\pm},\;\tilde{a}_m^{h\pm})^T$.

When substituting Eq.~(\ref{eq:source-field}) into Eq.~(\ref{eq:amp-src-approx}) one should evaluate the sources, and hence the local field amplitudes, at the layer center $x_3^c$. The resulting equation on the unknown amplitude vector can be written via some matrix operator $\Phi_{mn} (x_3^c)$ (to be explicitly given below) as
\begin{equation}
	\left( \begin{array}{c}\tilde{\bm{a}}^{+}_m(x_3^{(2)}) \\  \tilde{\bm{a}}^{-}_m(x_3^{(1)}) \end{array} \right) = \left( \begin{array}{c}\tilde{\bm{a}}^{ext,+}_m(x_3^{(2)}) \\ \tilde{\bm{a}}^{ext,-}_m(x_3^{(1)}) \end{array} \right) + \Delta x_3 \Phi_{mn}(x_3^c) \left( \begin{array}{c}\tilde{\bm{a}}^{+}_n(x_3^c) \\ \tilde{\bm{a}}^{-}_n(x_3^c) \end{array} \right) + O((\Delta x_3)^2).
	\label{eq:solution_1}
\end{equation}
One can find the unknown vector in the right-hand part by writing out an equation analogous to Eq.~(\ref{eq:amp-src-approx}) when both pairs of Eqs.~(\ref{eq:amp-src}) are evaluated at $x_3 = x_3^c$:
\begin{equation}
	\left( \begin{array}{c}\tilde{\bm{a}}^{+}_m(x_3^c) \\  \tilde{\bm{a}}^{-}_m(x_3^c) \end{array} \right) = \left( \begin{array}{c}\tilde{\bm{a}}^{ext,+}_m(x_3^c) \\ \tilde{\bm{a}}^{ext,-}_m(x_3^c) \end{array} \right) + \Delta x_3 \hat{\Phi}_{mn}(x_3^c) \left( \begin{array}{c}\tilde{\bm{a}}^{+}_n(x_3^c) \\ \tilde{\bm{a}}^{-}_n(x_3^c) \end{array} \right) + O((\Delta x_3)^2),
	\label{eq:solution_2}
\end{equation}
and the latter self-consistent linear algebraic equation is solved neglecting the $O((\Delta x_3)^2)$ terms. Operator $\hat{\Phi}_{mn}$ here differs from $\Phi_{mn}$ only by the exponential factor $\exp(ik_{3m}\Delta x_3/2)$. However, direct substitution of Eq.~(\ref{eq:solution_2}) into Eq.~(\ref{eq:solution_1}) shows that the inversion would provide an excessive accuracy, and it is suffices to take only the zero-order term of Eq.~(\ref{eq:solution_2}) into account. Thus, the local field amplitudes in the right-hand part of Eq.~(\ref{eq:solution_1}) can be replaced by the external ones. These amplitudes evaluated at the layer center are related with the known amplitudes at the layer boundaries through propagation factors: $\tilde{\bm{a}}^{ext,+}(x_3^c) = \tilde{\bm{a}}^{ext,+}(x_3^{(1)})\exp(ik_{3n}\Delta x_3/2)$, and $\tilde{\bm{a}}^{ext,-}(x_3^c) = \tilde{\bm{a}}^{ext,-}(x_3^{(2)})\exp(ik_{3n}\Delta x_3/2)$. Then, Eq.~(\ref{eq:amp-src-approx}) transforms to the following approximate relation:
\begin{equation}
	\begin{split}
		\left( \begin{array}{c}\tilde{\bm{a}}^{+}_m(x_3^{(2)}) \\  \tilde{\bm{a}}^{-}_m(x_3^{(1)}) \end{array} \right) &= \left( \begin{array}{c}\tilde{\bm{a}}^{ext,+}_m(x_3^{(1)}) \\ \tilde{\bm{a}}^{ext,-}_m(x_3^{(2)}) \end{array} \right)\exp(ik_{3m}\Delta x_3) + \frac{ik_b\Delta x_3}{2}\frac{k_b}{k_{3m}}\exp(ik_{3m}\Delta x_3/2) \\
		& \times \sum_n \exp(ik_{3n}\Delta x_3/2) \left[ (\mathcal{V}^{T})_{m}\Omega_{Emn}\mathcal{V}_{n} + (\mathcal{W}^{T})_{m}\Omega_{Hmn}\mathcal{W}_{n} \right] \left( \begin{array}{c}\tilde{\bm{a}}^{ext,+}_n(x_3^{(1)}) \\ \tilde{\bm{a}}^{ext,-}_n(x_3^{(2)}) \end{array} \right),
	\end{split}
	\label{eq:amp-solution}
\end{equation}
where $\Omega_{E,Hmn}$ are components of the Fourier block-Toeplitz matrices obtained by the Fourier transform of corresponding matrices $\Omega_{E,H}$ evaluated at coordinate $x_3^c$. The amplitudes of the first term in the right-hand side were translated using propagation factor $\exp(ik_{3n}\Delta x_3)$ so as to get identical vectors of external amplitudes in both terms. The accuracy of the derived equation is similar to other Fourier methods as they treat grating structures within each thin slice as homogeneous along the vertical coordinate.

Eq.~(\ref{eq:amp-solution}) directly provides relations between incoming and outgoing wave amplitudes for diffraction on a thin grating slice. The above derivation supposes that external field amplitudes can be generated by any sources located outside the layer $x_3^{(1)}\leq x_3\leq x_3^{(2)}$. Since only amplitudes propagating toward the layer (incoming) are present in the right-hand part of Eq.~(\ref{eq:amp-solution}) we can associate them with local fields at boundaries of the layer and leave out the superscript ''ext''. By simple reordering of these equations one can compose an $S$-matrix for the corresponding layer. Given a planar structure bounded by planes $x_3=x_3^{low}$ and $x_3=x_3^{up}$, we will refer to the $S$-matrix of this structure as a $2\times2$ block matrix that relates plane wave amplitudes at the boundaries as follows:
\begin{equation}
	\left( \begin{array}{c}\tilde{\bm{a}}^{-}_m(x_3^{low}) \\ \tilde{\bm{a}}^{+}_m(x_3^{up}) \end{array} \right) = \sum_n \left(\begin{array}{cc} S_{11,mn}&S_{12,mn}\\S_{21,mn}&S_{22,mn} \end{array}\right) \left( \begin{array}{c}\tilde{\bm{a}}^{+}_m(x_3^{low}) \\ \tilde{\bm{a}}^{-}_m(x_3^{up}) \end{array} \right)
	\label{eq:S-matrix-22}
\end{equation}

To apply Eqs.~(\ref{eq:amp-solution}), (\ref{eq:S-matrix-22}) to a deep structure, the grating layer should be divided into a number of slices analogous to the Fourier Modal Method (FMM) and other Fourier methods. However, the calculation of eigen modes is no longer needed since $S$-matrix components for each slice are readily available. Once the $S$-matrix of each slice is attained via Eq.~(\ref{eq:amp-solution}) a corresponding matrix for the whole grating layer can be calculated by the $S$-matrix propagation algorithm known to be numerically stable \cite{Sambles1995,Li1996-1}. The algorithm is based on the following multiplication rule. Given two $S$-matrices $S^{(1,2)}$ of planar structures occupying adjacent plane layers $x_3^{(1)}\leq x_3\leq x_3^{(2)}$ and $x_3^{(2)}\leq x_3\leq x_3^{(3)}$ respectively, the components of the $S$-matrix that relates amplitudes at interfaces $x_3 = x_3^{(1)}$ and $x_3 = x_3^{(3)}$ are
\begin{equation}
	\begin{split}
	&S_{11} = S_{11}^{(1)} + S_{12}^{(1)}\left(I-S_{11}^{(2)}S_{22}^{(1)}\right)^{-1}S_{11}^{(2)}S_{21}^{(1)}, \\
	&S_{12} = S_{12}^{(1)}\left(I-S_{11}^{(2)}S_{22}^{(1)}\right)^{-1}S_{12}^{(2)}, \\
	&S_{21} = S_{21}^{(2)}\left(I-S_{22}^{(1)}S_{11}^{(2)}\right)^{-1}S_{21}^{(1)}, \\
	&S_{22} = S_{22}^{(2)} + S_{21}^{(2)}\left(I-S_{22}^{(1)}S_{11}^{(2)}\right)^{-1}S_{22}^{(1)}S_{12}^{(2)}.
	\end{split}
	\label{eq:S-miltiplication}
\end{equation}
Due to the presence of matrix inversions in Eq.~(\ref{eq:S-miltiplication}) the complexity of the algorithm can be estimated as $O(N^3_FN_S)$, where $N_F$ is a number of Fourier harmonics, and $N_S$ is a number of slices.


\section{Examples}

Matrices $\Omega_{E,H}$ generally depend on a particular implementation of the Fourier approach and follow from numerous results attained by different authors on the Fourier Modal Method, Differential Method, C-method, and the Generalized Source Method together with other volume integral implementations (e.g., \cite{Li1996,Popov2001,Granet2002,Schuster2007,Beurden2017}). For the reader to get acquainted with possible implementations of the previous section we provide here two illustrative examples of widely used gratings followed by a derivation of an S-matrix for sinusoidally corrugated layers of 2D materials.

\subsection{1D lamellar grating}

A 1D lamellar grating is one of the simplest but nevertheless among the most practically important examples of periodic corrugations. Consider a formulation in Cartesian coordinates. Then, all components $\Omega_H$ would be zero, and the matrix $\Omega_E$ would be diagonal. Correct factorization of the products of discontinuous functions \cite{Granet1996,Lalanne1996,Li1996,Popov2001} yields
\begin{equation}
	\Omega_{Emn} = \left( \begin{array}{ccc} [\varepsilon_b/\varepsilon]_{mn}^{-1} - \delta_{mn}&&\\&[\varepsilon/\varepsilon_b]_{mn} - \delta_{mn}& \\ &&\delta_{mn}-[\varepsilon/\varepsilon_b]^{-1}_{mn} \end{array} \right).
	\label{eq:omega-lam}
\end{equation}
Here, periodicity along $x_1$ is assumed, and $[\varepsilon/\varepsilon_b]_{mn}^{-1}$, $[\varepsilon_b/\varepsilon]_{mn}^{-1}$ denote inverted truncated Fourier-matrices of periodic permittivity functions. The permittivity function in the considered case explicitly writes $\varepsilon(x_1)=\varepsilon_1$, $(k-\alpha/2)\Lambda\leq x_1<(k+\alpha/2)\Lambda$, and $\varepsilon(x_1)=\varepsilon_2$, $(k+\alpha/2)\Lambda\leq x_1<(k+1-\alpha/2)\Lambda$ with $k\in\mathbb{Z}$, $0<\alpha<1$, and $\varepsilon_{1,2}=const$, and has the following elements of the Fourier matrix:
$$[\varepsilon/\varepsilon_b]_{mn}=\frac{\varepsilon_2}{\varepsilon_b}\delta_{m,n} + \alpha\frac{\varepsilon_1-\varepsilon_2}{\varepsilon_b}\sinc[\pi\alpha(m-n)]$$
In the collinear diffraction case when $k_2^{(0)}=0$ the $S$-matrix splits into two independent parts for the TE and TM polarization separately. We get for the TE polarization 
\begin{equation}
	\begin{split}
	&S_{11mn}^{(e)} = S_{22mn}^{(e)} = \frac{ik_b\Delta x_3}{2}\frac{k_b}{k_{3m}} \exp[i(k_{3m}+k_{3n})\Delta x_3/2]\left( [\varepsilon/\varepsilon_b]_{mn} - \delta_{mn} \right), \\
	&S_{12mn}^{(e)} = S_{21mn}^{(e)} = \exp(ik_{3m}\Delta x_3)+ \frac{ik_b\Delta x_3}{2}\frac{k_b}{k_{3m}} \exp[i(k_{3m}+k_{3n})\Delta x_3/2]\left( [\varepsilon/\varepsilon_b]_{mn} - \delta_{mn} \right),
	\end{split}
	\label{eq:S-lam-TE}
\end{equation}
and for the TM  polarization
\begin{subequations}
\begin{equation}
	\begin{split}
	S_{11mn}^{(h)} = S_{22mn}^{(h)} = &\frac{ik_b\Delta x_3}{2}\frac{k_b}{k_{3m}} \exp[i(k_{3m}+k_{3n})\Delta x_3/2] \\
	&\times \left\{ - \frac{k_{3m}}{k_b} \left([\varepsilon_b/\varepsilon]_{mn}^{-1}-\delta_{mn}\right)\frac{k_{3n}}{k_b} + \frac{k_{1m}}{k_b} \left(\delta_{mn}-[\varepsilon/\varepsilon_b]^{-1}_{mn}\right)\frac{k_{1n}}{k_b} \right\}
	\end{split}
\end{equation}
\begin{equation}
	\begin{split}
	S_{12mn}^{(h)} = S_{21mn}^{(h)} = &\exp(ik_{3m}\Delta x_3)+ \frac{ik_b\Delta x_3}{2}\frac{k_b}{k_{3m}} \exp[i(k_{3m}+k_{3n})\Delta x_3/2] \\
	&\times \left\{ \frac{k_{3m}}{k_b} \left([\varepsilon_b/\varepsilon]_{mn}^{-1}-\delta_{mn}\right)\frac{k_{3n}}{k_b} + \frac{k_{1m}}{k_b} \left(\delta_{mn}-[\varepsilon/\varepsilon_b]^{-1}_{mn}\right)\frac{k_{1n}}{k_b} \right\}
	\end{split}
\end{equation}
	\label{eq:S-lam-TM}
\end{subequations}

Independence of function $\varepsilon({\bf r})$ of coordinate $x_3$ allows calculating the grating $S$-matrix with a reduced complexity since in this case the expressions in Eqs.~(\ref{eq:S-lam-TE}), (\ref{eq:S-lam-TM}) should be evaluated only once for a single thin layer. Suppose one starts with S-matrix $S^{(0)}$ of a layer with thickness $\Delta x_3$. S-matrix $S^{(k)}$ of a layer having thickness $2^k\Delta x_3$ is a composition of matrices for half-depth layers: $S^{(k)} = S^{(k-1)}*S^{(k-1)}$. Thus, for $x_3$-invariant corrugations the complexity of the method reduces down to $O(N_F^3\log N_S)$.

\begin{figure}[ht!]
\centering\includegraphics[width=10cm]{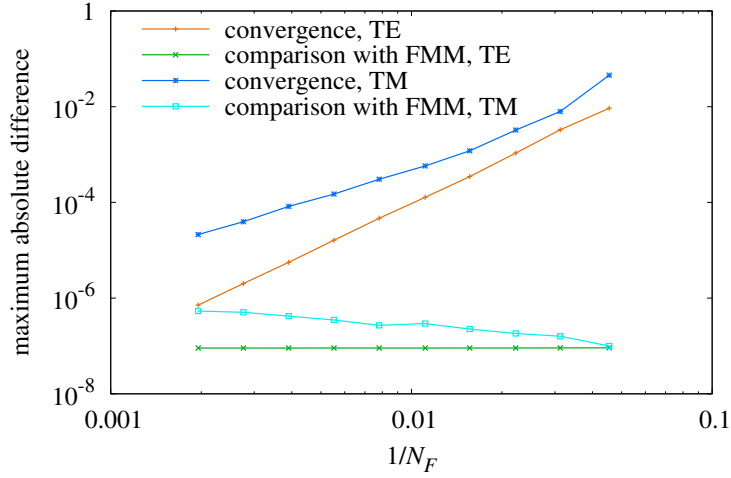}
\caption{The convergence of the method based on the direct S-matrix calculation, Eqs.~(\ref{eq:S-lam-TE}), (\ref{eq:S-lam-TM}), and the comparison with S-matrices calculated by the Fourier Modal Method. The maximum absolute difference between corresponding S-matrix components is plotted against the inverse number of Fourier harmonics. The grating parameters are: period-to-wavelength ratio $\Lambda/\lambda = 1.5$, depth-to-wavelength ratio $h/\lambda = 0.5$, and $\alpha = 0.5$. Substrate and cover refractive indices are $1.5$ and $1$ respectively. Grating permittivity $\varepsilon_1 = 6.25$.}
\label{fig:conv_d}
\end{figure}

\begin{figure}[ht!]
\centering\includegraphics[width=10cm]{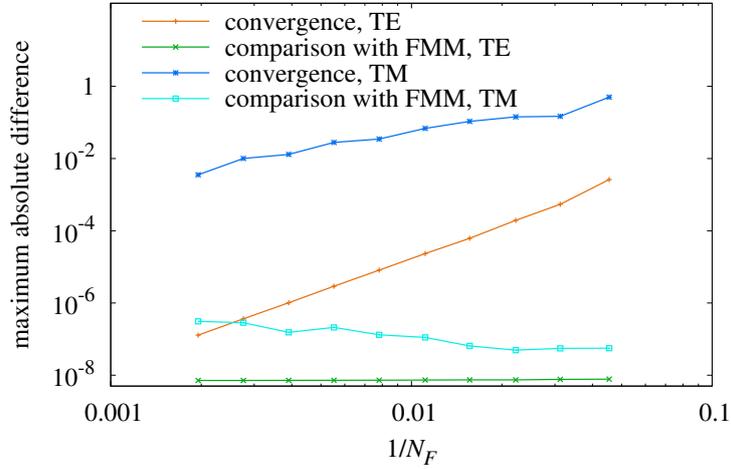}
\caption{Same as in Fig.~\ref{fig:conv_d}, but for metallic grating of permittivity $\varepsilon_1 = -9.6+1.1i$.}
\label{fig:conv_m}
\end{figure}

Figures \ref{fig:conv_d} and \ref{fig:conv_m} demonstrate the convergence of the developed S-matrix method and show the comparison with solutions obtained by the FMM as it is described in \cite{Gushchin2010} for lamellar dielectric and metallic gratings respectively. The maximum absolute difference in corresponding S-matrix components is plotted along vertical axis. Grating parameters used in the calculations are indicated in the figure captions. Difference of the obtained solutions from the FMM results is seen to be much below the accuracy of both methods, and depends on initial layer depth $\Delta x_3$ (see the previous paragraph), which was taken to be $2^{-30}h$. The FMM is known to be particularly efficient for gratings with vertical walls, so the calculation time of the direct S-matrix method in the presented example was about $10$ times larger than that of the FMM for each value of $N_F$, though this time difference can be reduced by increasing the value of $\Delta x_3$ as long as the reduction of the accuracy can be tolerated.

\subsection{1D sinusoidal grating in curvilinear coordinates}

\begin{figure}[ht!]
\centering\includegraphics[width=10cm]{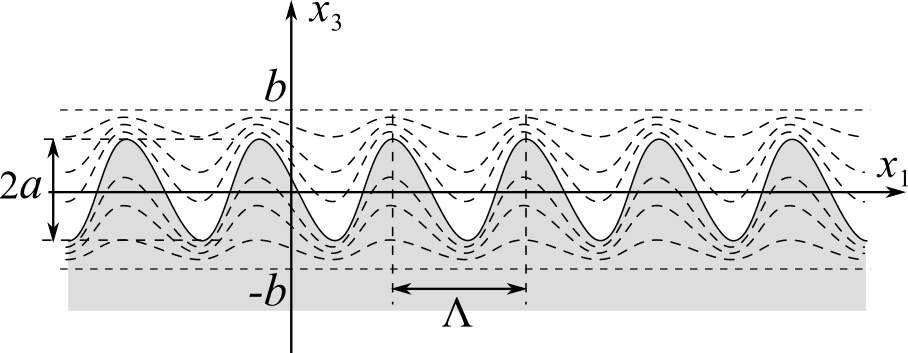}
\caption{Illustration of the Curvilinear Coordinate Generalized Source Method idea: Maxwell's equations write in a curvilinear coordinate system $(z^1,z^2,z^3)$ such that one of its coordinate planes exactly fits corrugation profile, and these coordinates continuously become Cartesian $(x_1,x_2,x_3)$ in the region in the vicinity of the grating. Dashed lines demonstrate  coordinate planes $z^3=const$.}
\label{fig:gsmcc}
\end{figure}

The necessity of the magnetic sources in the above derivations is due to the previously developed curvilinear coordinate Generalized Source Method with effective Generalized Metric Sources \cite{Shcherbakov2013,Shcherbakov2017}. The core idea is to fit a grating corrugation profile with a suitable cuvilinear coordinate transformation $(x_1,x_2,x_3)\rightarrow(z^1,z^2,z^3)$ similar to what is done in the C-method \cite{Chandezon1980}, while the chosen curvilinear coordinates should continuously become Cartesian in a region near the grating. Fig.~\ref{fig:gsmcc} demonstrates an example of a periodic corrugation with coordinate planes of a new system. This approach exploits the similarity between Maxwell's operators in the Cartesian and the curvilinear metric (see e.g. Chapter 8 of \cite{Popov2012}), and a possibility to treat metric contributions as source terms \cite{Shcherbakov2013}. Under this rationale the Generalized Sources derived from local metric variations read
\begin{equation}
	\begin{split}
	&J^{\alpha} = -i\omega\varepsilon_b\left( \frac{\varepsilon}{\varepsilon_b} \sqrt{g} g^{\alpha\beta} - \delta^{\alpha\beta} \right) E_{\beta}, \\
	&M^{\alpha} = -i\omega\mu_b\left( \frac{\mu}{\mu_b} \sqrt{g} g^{\alpha\beta} - \delta^{\alpha\beta} \right) H_{\beta},
	\end{split}
	\label{eq:JM-curv}
\end{equation}
where $g^{\alpha\beta}$ are local metric tensor components, $g=1/\det\{g^{\alpha\beta}\}$, and summation over the repeated indices is implied. These sources can be directly substituted into Eqs.~(\ref{eq:E_volume}) with functions Eqs.~(\ref{eq:Green_e}), (\ref{eq:Green_m}) under direct replacement of Cartesian coordinates $(x_1,x_2,x_3)$ with curvilinear ones $(z^1,z^2,z^3)$. The approach yields matrices $\Omega$ that depend on local metric and medium properties \cite{Shcherbakov2013}:
\begin{equation}
	\Omega_{E,Hmn} = \left( \begin{array}{ccc} \left[\eta_{E,H}/(\sqrt{g}g^{33})\right]_{mn}-\delta_{mn}&0&\left[g^{13}/g^{33}\right]_{mn} \\ 0&[\eta_{E,H}\sqrt{g}]_{mn}-\delta_{mn}&0 \\ \left[g^{13}/g^{33}\right]_{mn}&0&\delta_{mn}-\left[1/(\eta_{E,H}\sqrt{g}g^{33})\right]_{mn} \end{array} \right).
	\label{eq:omega-sin}
\end{equation}
with $\eta_E=\varepsilon/\varepsilon_b$, $\eta_H = \mu/\mu_b$. The latter two fractions are constant within each slice in curvilinear coordinates, and can be taken out of the Fourier matrices. $S$-matrices for the two polarizations then explicitly read:
\begin{subequations}
\begin{equation}
	\begin{split}
	S^{(e,h)}_{11mn} = &\frac{ik_b\Delta z^3}{2}\frac{k_b}{k_{3m}} \exp[i(k_{3m}+k_{3n})\Delta z^3/2] \bigg\{ \eta_{E,H}[\sqrt{g}]_{mn} - \delta_{mn} \\
	&- \frac{k_{3m}}{k_b} \left( \eta_{H,E} \left[\frac{1}{\sqrt{g}g^{33}}\right]_{mn}-\delta_{mn}\right)\frac{k_{3n}}{k_b} + \frac{k_{1m}}{k_b} \left(\delta_{mn}-\frac{1}{\eta_{H,E}}\left[\frac{1}{\sqrt{g}g^{33}}\right]_{mn}\right)\frac{k_{1n}}{k_b} \\
	&+ \frac{k_{3m}}{k_b}\left[\frac{g^{13}}{g^{33}}\right]_{mn}\frac{k_{1n}}{k_b} - \frac{k_{1m}}{k_b}\left[\frac{g^{31}}{g^{33}}\right]_{mn}\frac{k_{3n}}{k_b} \bigg\}
	\end{split}
\end{equation}
\begin{equation}
	\begin{split}
	S^{(e,h)}_{12mn} = &\exp(ik_{3m}\Delta z^3) + \frac{ik_b\Delta z^3}{2}\frac{k_b}{k_{3m}} \exp[i(k_{3m}+k_{3n})\Delta z^3/2] \bigg\{ \eta_{E,H}[\sqrt{g}]_{mn} - \delta_{mn} \\
	&+ \frac{k_{3m}}{k_b} \left( \eta_{H,E} \left[\frac{1}{\sqrt{g}g^{33}}\right]_{mn}-\delta_{mn}\right)\frac{k_{3n}}{k_b} + \frac{k_{1m}}{k_b} \left(\delta_{mn}-\frac{1}{\eta_{H,E}}\left[\frac{1}{\sqrt{g}g^{33}}\right]_{mn}\right)\frac{k_{1n}}{k_b} \\
	&- \frac{k_{3m}}{k_b}\left[\frac{g^{13}}{g^{33}}\right]_{mn}\frac{k_{1n}}{k_b} - \frac{k_{1m}}{k_b}\left[\frac{g^{31}}{g^{33}}\right]_{mn}\frac{k_{3n}}{k_b} \bigg\}
	\end{split}
\end{equation}
\begin{equation}
	\begin{split}
	S^{(e,h)}_{21mn} = &\exp(ik_{3m}\Delta z^3) + \frac{ik_b\Delta z^3}{2}\frac{k_b}{k_{3m}} \exp[i(k_{3m}+k_{3n})\Delta z^3/2] \bigg\{ \eta_{E,H}[\sqrt{g}]_{mn} - \delta_{mn} \\
	&+ \frac{k_{3m}}{k_b} \left( \eta_{H,E} \left[\frac{1}{\sqrt{g}g^{33}}\right]_{mn}-\delta_{mn}\right)\frac{k_{3n}}{k_b} + \frac{k_{1m}}{k_b} \left(\delta_{mn}-\frac{1}{\eta_{H,E}}\left[\frac{1}{\sqrt{g}g^{33}}\right]_{mn}\right)\frac{k_{1n}}{k_b} \\
	&+ \frac{k_{3m}}{k_b}\left[\frac{g^{13}}{g^{33}}\right]_{mn}\frac{k_{1n}}{k_b} + \frac{k_{1m}}{k_b}\left[\frac{g^{31}}{g^{33}}\right]_{mn}\frac{k_{3n}}{k_b} \bigg\}
	\end{split}
\end{equation}
\begin{equation}
	\begin{split}
	S^{(e,h)}_{22mn} = &\frac{ik_b\Delta z^3}{2}\frac{k_b}{k_{3m}} \exp[i(k_{3m}+k_{3n})\Delta z^3/2] \bigg\{ \eta_{E,H}[\sqrt{g}]_{mn} - \delta_{mn} \\
	&- \frac{k_{3m}}{k_b} \left( \eta_{H,E} \left[\frac{1}{\sqrt{g}g^{33}}\right]_{mn}-\delta_{mn}\right)\frac{k_{3n}}{k_b} + \frac{k_{1m}}{k_b} \left(\delta_{mn}-\frac{1}{\eta_{H,E}}\left[\frac{1}{\sqrt{g}g^{33}}\right]_{mn}\right)\frac{k_{1n}}{k_b} \\
	&- \frac{k_{3m}}{k_b}\left[\frac{g^{13}}{g^{33}}\right]_{mn}\frac{k_{1n}}{k_b} + \frac{k_{1m}}{k_b}\left[\frac{g^{31}}{g^{33}}\right]_{mn}\frac{k_{3n}}{k_b} \bigg\}
	\end{split}
\end{equation}
\label{eq:S-sin}
\end{subequations}
These equations have a symmetric form relative to a TE/TM polarization change contrary to the previous example, which is a consequence of the symmetry of the Generalized Metric sources in Eq.~(\ref{eq:JM-curv}). In case of a sinusoidal corrugation the coordinate transformation is defined as $z^{1,2}=x_{1,2}$, and $x_3=z^3 + (1-|z^3|/b)a\sin(Kx_1)$, where $h=2a$ is the corrugation depth, and $-b\leq x_3\leq b,\;b\geq a$ is a region with curvilinear metric (see Fig.~\ref{fig:gsmcc} for illustration and \cite{Shcherbakov2013} for details). The components of Fourier matrices for sinusoidally corrugated gratings are found analytically:
\begin{equation}
	\begin{split}
	[\sqrt{g}]_{mn} &= \delta_{m,n} - \frac{1}{2}\alpha\sign(z^3)(\delta_{m,n+1}+\delta_{m,n-1}) \\
	\left[\frac{1}{\sqrt{g}g^{33}}\right]_{mn} &= \left(\frac{\sqrt{1+2\chi^2}-1}{\sqrt{2}\chi}\right)^{2k} \left[ \frac{\delta_{m-n,2k}}{\sqrt{1+2\chi^2}} - \alpha\sign(z^3)\delta_{m-n,2k+1}\frac{\sqrt{1+2\chi^2}-1}{2\chi^2} \right]  \\	
	\left[\frac{g^{13}}{g^{33}}\right]_{mn} &= \frac{i}{\sqrt{2}\chi} \left(\frac{\sqrt{1+2\chi^2}-1}{\sqrt{2}\chi}\right)^{2k} \left[  \alpha\sign(z^3)\delta_{m-n,2k} - \delta_{m-n,2k+1}\frac{\sqrt{1+2\chi^2}-1}{\sqrt{1+2\chi^2}} \right]
	\end{split}
	\label{eq:g-sin}
\end{equation}
with $\alpha=a/b$ and $\chi = (1-|z^3|/b)^2(\alpha Kh)^2/2$.

The two considered examples demonstrate the possibility to derive $S$-matrix components explicitly, though the resulting expressions can be rather bulky. Therefore, we restrict ourselves here to these two examples for gratings of bulk materials as $S$-matrix components in other various cases can be attained with aids of the vast literature on the Fourier methods and general equations of the previous section. The method has a second order polynomial convergence with regard to the slice number, and typical convergence plots are quite similar to those presented for the GSM in \cite{Shcherbakov2012,Shcherbakov2013,Shcherbakov2017}. The convergence rate relative to the number of Fourier orders for profiled gratings and S-matrices written in Cartesian coordinates is polynomial, and for continuously differentiable profiles with curvilinear coordinate S-matrices is exponential, again, similarly to the GSM \cite{Shcherbakov2012}, and the GSMCC \cite{Shcherbakov2013} respectively.

\subsection{Corrugated 2D material}

We proceed by modifying the results of the previous subsection and consider a layer of 2D material, e.g., graphene, on top of a corrugated substrate. Here we focus on 1D holographic gratings, whose profiles can be well approximated by sinusoidal functions introduced above.

Electric currents in such materials depend only on tangential electric field components. Therefore, within the rationale of a curvilinear coordinate transformation, the relation between the current and the field via surface conductivity $\sigma_s$ writes as follows
\begin{equation}
	J^{\alpha} = \sigma_s\delta(z^3)\frac{\sqrt{g}}{g_{\alpha\alpha}}({\bf E}\cdot\bm{e}_{\alpha}) = \sigma_s\frac{\sqrt{g}}{g_{\alpha\alpha}}\delta(z^3) E_{\alpha},\;\alpha = 1,2,
	\label{eq:JsE-curv}
\end{equation}
where upper and lower indices distinguish contravariant and covariant vector components, and $\bm{e}_{\alpha} = (\partial x_{\beta}/\partial z^{\alpha})\hat{\bm{e}}_{\beta}$ are tangent vectors to curvilinear coordinate planes. Here we suppose the 2D material layer to coinside with surface $z^3=0$. It is seen that the normalization factor $\sqrt{g}/g_{\alpha\alpha}$ makes effective conductivity in the curvilinear metric be periodic even when the effect of corrugation on conductivity is neglected. A possible impact of the periodicity on $\sigma_s$ \cite{Jonson2008} can be also included in the present method straightforwardly, though we do not account for it in the following examples.

Substitution of sources given by Eq.~(\ref{eq:JsE-curv}) into Eq.~(\ref{eq:amp-src}) yields relations
\begin{equation}
	\begin{split}
	\tilde{a}_m^{e\pm}(\pm0) &= \tilde{a}_m^{ext,e\pm}(\pm0) - \frac{\omega\mu_b}{2k_{m3}} \hat{\bm{e}}_m^{e\pm}\cdot {\bf J}_m(0) \\ 
	\tilde{a}_m^{h\pm}(\pm0) &= \tilde{a}_m^{ext,h\pm}(\pm0) - \frac{\omega\mu_b}{2k_{m3}} \hat{\bm{e}}_m^{h\pm}\cdot {\bf J}_m(0) \\ 
	\end{split}
	\label{eq:aJs}
\end{equation}
There are two differences between the latter equations and Eqs.~(\ref{eq:amp-src-approx}): due to presence of delta-function in Eq.~(\ref{eq:JsE-curv}) Eqs.~(\ref{eq:aJs}) are exact, and terms proportional to layer thickness $\Delta z^3$ with curvilinear sources of Eq.~(\ref{eq:JM-curv}) are absent. The external field is continuous across the $z^3=0$ plane, so $\tilde{a}_m^{ext,\pm}(+0) = \tilde{a}_m^{ext,\pm}(-0)$ for both polarizations. The electric current at layer location ${\bf J}_m(0)$ depends on the continuous tangential electric field. In case of a 1D corrugation along $x_1$ direction and collinear diffraction with $k_{m2}=0$ this tangential field reads $\tilde{E}_{m2}(0) = \tilde{a}_m^{e+}(\pm0) + \tilde{a}_m^{e-}(\pm0)$ for the TE polarization, and $\tilde{E}_{m1}(0) = (k_{3m}/k_b)\left[\tilde{a}_m^{h-}(\pm0) - \tilde{a}_m^{h+}(\pm0)\right]$ for the TM polarization. Note that in this case the field which contributes to the source cannot be substituted by the external field as was the case in Eq.~(\ref{eq:amp-solution}) derivation, and a self-consistent equation system should be solved. This yields the following S-matrix relations for the two polarizations:
\begin{equation}
	\left( \begin{array}{c} \tilde{a}_m^{e-}(-0)\\\tilde{a}_m^{e+}(+0) \end{array} \right) = \frac{1}{1+\zeta^e_m} \left( \begin{array}{cc} -\zeta^e_m & 1 \\ 1 & -\zeta^e_m \end{array} \right) \left( \begin{array}{c} \tilde{a}_m^{e+}(-0)\\\tilde{a}_m^{e-}(+0) \end{array} \right)
	\label{eq:S-2D-TE}
\end{equation}
\begin{equation}
	\begin{split}
	\left( \begin{array}{c} \tilde{a}_m^{h-}(-0)\\\tilde{a}_m^{h+}(+0) \end{array} \right) &= \left[ \delta_{mn} + (\sqrt{g}/g_{11})_{mn}\zeta^h_n \right]^{-1} \\ &\times \left( \begin{array}{cc} (\sqrt{g}/g_{11})_{mn}\zeta_n^h & \delta_{mn} \\ \delta_{mn} & (\sqrt{g}/g_{11})_{mn}\zeta_n^h \end{array} \right) \left( \begin{array}{c} \tilde{a}_n^{h+}(-0)\\\tilde{a}_n^{h-}(+0) \end{array} \right),
	\end{split}
	\label{eq:S-2D-TM}
\end{equation}
where $\zeta^e_m = \sigma_s\omega\mu_b/2k_{m3}$, $\zeta^h_m = \sigma_s k_{m3}/2\omega\varepsilon_b$. Inversion implies that the corresponding matrix should be composed of the truncated Fourier matrix $\{\sqrt{g}/g_{11}\}_{m,n=1}^{N}$ and then numerically inverted. Eq.~(\ref{eq:S-2D-TE}) consists of known plane wave reflection and transmission coefficients for the TE polarization \cite{Geim2008}. Similarly Eq.~(\ref{eq:S-2D-TM}) for the TM polarization would reduce to such coefficients in the absence of the corrugation, when $\sqrt{g}=g_{11}=1$. It is important to note that Eqs.~(\ref{eq:S-2D-TE}), (\ref{eq:S-2D-TM}) should be applied together with the results of the previous subsection.


\section{Resonant reflection by corrugated graphene}

The derived S-matrices for 1D holographic gratings, Eqs.~(\ref{eq:S-sin}), and a corrugated 2D material layer, Eq.~(\ref{eq:S-2D-TE}),(\ref{eq:S-2D-TM}) can be used to simulate the optical response of graphene sheets covering periodically structured substrates of arbitrary corrugation depth. The present method is superior to approaches used in \cite{Bludov2013,Nikitin2013} since it doest not rely on the Rayleigh hypothesis, and consequently does hot suffer from corresponding convergence issues inherent to Rayleigh methods.

Consider a graphene monolayer on top of a corrugated Si substrate at room temperature. Silicon permittivity around 10~THz can be taken to be constant being approximately equal to 11.5. Graphene dispersion at room temperature \cite{Falkovsky2007} is described by equation
\begin{equation}
	\frac{\sigma(\omega)}{\sigma_0} = \frac{8i}{\pi}\frac{T}{\omega+i\tau^{-1}}\log\left[2\cosh\left(\frac{E_F}{2T}\right)\right] + H\left(\frac{\omega}{2}\right) + \frac{4i\omega}{\pi}\int\limits_0^{\infty} d\epsilon\frac{H(\epsilon)-H(\omega/2)}{\omega^2-4\epsilon^2},
	\label{eq:graphene}
\end{equation}
where $\sigma_0=e^2/4\hbar$, and $H(\epsilon) = \sinh(\epsilon/T)/\left[ \cosh(E_F/T)+\cosh(\epsilon/T) \right]$. Here, the Fermi-energy depends on the applied gate voltage, and can be tuned in a wide range. Relaxation rate $\tau^{-1}$ depends both on the quality of graphene (which in turn depends on the fabrication process), the quality of the substrate and presence of a BN interlayer, so that $\tau$ can vary by more than an order of magnitude $4\cdot10^{-14}$~s~$\lesssim\tau\lesssim 10^{-12}$~s (e.g., see experimental results in \cite{Ruoff2009,Dean2010,Mayorov2011}). For further examples we assume $\tau = 10^{-13}$~\textit{s}, and $E_F = 0.4$~\textit{eV}.

Graphene sheets are known to support highly confined surface plasmon-polaritons in the terahertz band. Condition
\begin{equation}
	f(\omega,k_1) = \frac{\sigma(\omega)}{\omega} + \frac{\varepsilon_1(\omega)}{\sqrt{\omega^2\varepsilon_1(\omega)\mu_0 - k_1^2}} + \frac{\varepsilon_2(\omega)}{\sqrt{\omega^2\varepsilon_2(\omega)\mu_0 - k_1^2}} = 0
	\label{eq:spp-dispersion}
\end{equation}
defines the dispersion of the TM surface plasmons in graphene \cite{Bludov2013}. Here $\varepsilon_{1,2}$ are permittivities of the homogeneous media below and above the 2D sheet, and $k_1$ is the projection of the wavevector on the sheet plane. The condition $k_1\gg \Re(\omega^2\varepsilon\mu_0)$ provides a good analytical approximation in case of low absorbing substrates. In case of a periodically corrugated sheet a surface plasmon wave should be coupled with an evanescent grating order to attain a resonance condition \cite{Avrutsky2000}.

\begin{figure}
\centering\includegraphics[width=10cm]{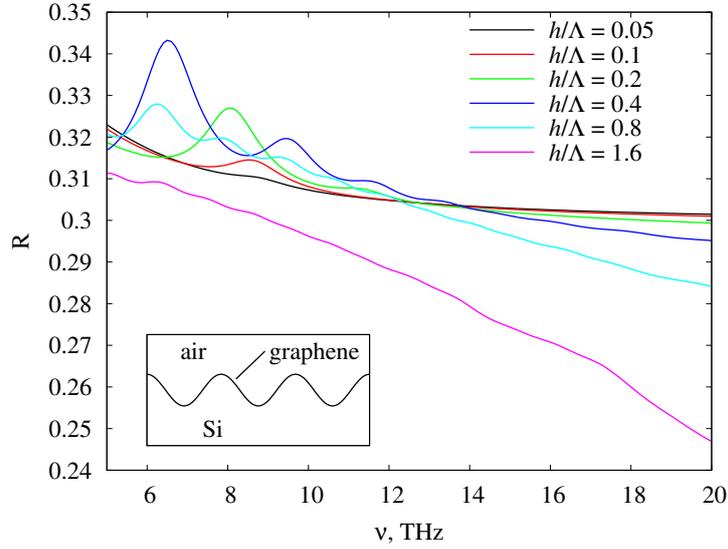}
\caption{Dependence of zero order reflection efficiency from the frequency for the normally incident plane wave for corrugated graphene sheet placed on top of silicon substrate with corrugation period $0.8\,\mu m$ and varying corrugation depth.}
\label{fig:res_depth}
\end{figure}

Large values of $k_1$ require the use of small-period gratings. Fig.~\ref{fig:res_depth} demonstrates reflection resonance for a sinusoidally corrugated silica substrate with a graphene sheet on top of it with grating period $\Lambda = 0.8$~$\mu m$ and varying corrugation depth. Normal incidence is chosen. A single feasible resonance peak for small ratios $h/\Lambda$ corresponds to the excitation of the surface plasmon-polariton by the first grating order. When the corrugation depth-to-period ratio increases this peak redshifts (similar behaviour can be seen in Fig.~5 of \cite{Nikitin2013}). The reason is the decrease of the local curvature radius which affects the surface plasmon dispersion, as Eq.~(\ref{eq:spp-dispersion}) is no more rigorous for corrugated layers. Interestingly, this is opposite to the blueshift of surface plasmon-polaritons supported by curved metal-dielectric interfaces \cite{Kim2016}. In addition to the redshift secondary resonant peaks become significant. These peaks are due to plasmon coupling with higher grating orders. Further increase of the depth-to-period ratio results in the  disappearance of significant resonances within the investigated frequency band.

\begin{figure}
\centering\includegraphics[width=10cm]{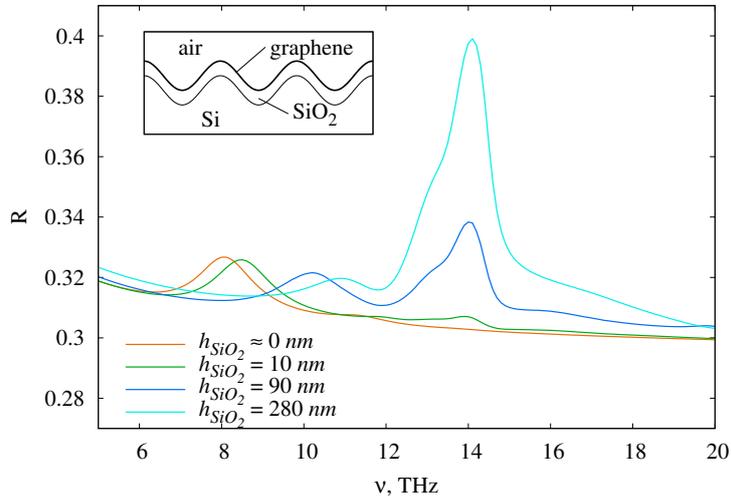}
\caption{Similar to Fig.~\ref{fig:res_depth}, but the graphene sheet is supposed to be in contact with an intermediate layer silica layer of varying thickness $h_{SiO_2}$.}
\label{fig:res_thk}
\end{figure}

In experiments silicon substrates are always covered by a layer of silicon dioxide whose thickness can vary from few to hundreds of nanometers. SiO$_2$ is generally a bad material for the terahertz band due to high resonant absorption (see measurements of amorphous silica permittivity in \cite{Popova1972}). The dispersion obtained in \cite{Popova1972} can be used to estimate the influence of silica layers on the resonant curves demonstrated in Fig.~\ref{fig:res_depth}. Fig.~\ref{fig:res_thk} shows such influence for a grating with $h/\Lambda = 0.2$ and different thickness $h_{SiO_2}$ of the silica interlayer. Few nanometer thick layers slighly blueshift the resonance peak, whereas for a $h_{SiO_2}\sim100$~\textit{nm} resonant absorption in silica itself around $14$~\textit{THz} prominently affects the reflection curve.

\section{Conclusion}

To sum up we derived a general form of S-matrix in the Fourier basis Eq.~(\ref{eq:amp-solution}) from the volume integral solution of Maxwell's equations. The provided examples demonstrate the possibility to attain explicit analytical S-matrix components, as shown for 1D lamellar and sinusoidally corrugated gratings of bulk materials. S-matrices for other types of gratings including complex 2D periodic diffractive optical elements and metasurfaces can be further derived in a similar manner on the basis of all previous results for the Fourier Modal Method, and other Fourier space methods (e.g., \cite{Popov2001,Granet2002,Schuster2007,Beurden2017}), and there is no need for solving the eigenvalue problem in each slice \cite{Sipe2008}. Moreover, the derived analytical S-matrix components can be directly used for resonant analysis of grating structures proposed in \cite{Bykov2013}. The obtained S-matrix components have second order accuracy relative to spatial discretization of the grating layer, which is similar to other Fourier space methods \cite{Popov2002}. In addition we expressed the S-matrix of a corrugated layer of 2D material with a given conductivity, which can be useful for further research of optical response functions of metasurfaces covered with such materials and multilayer quasi-2D structures.



\section*{Acknowledgements}

The work was supported by the Russian Scientific Foundation, Grant No. 17-79-20345.

\section*{Appendix}
This Appendix provides a sketch of the derivation of Eqs.~(\ref{eq:Green_e}), (\ref{eq:Green_m}). Due to the linearity of excitations Helmholtz Eqs.~(\ref{eq:Helmholtz}) can be considered separately either for the electric or magnetic sources. Suppose ${\bf M}=0$, and make a passage to scalar and vector potentials $\varphi$ and $\bf A$:
\begin{equation}
	{\bf E} = -\nabla\varphi + i\omega{\bf A},\;{\bf H} = (1/\mu_b) \nabla\times{\bf A}.
	\label{eq:potentials}
\end{equation}
Under Lorentz gauge condition $\omega\varepsilon_b\mu_b\varphi + i\nabla{\bf A}=0$ Maxwell's equations reduce to the Helmholtz equation on the vector potential
\begin{equation*}
	\Delta{\bf A} + k_b^2{\bf A} = -\mu_b{\bf J},
\end{equation*}
whose solution writes via scalar Green's function $g_0$
\begin{equation}
	{\bf A} = \mu_b \int g_0(\bm{r}-\bm{r}'){\bf J}(\bm{r}')d^3\bm{r}' = \mu_b \int \frac{\exp(ik_b|\bm{r}-\bm{r}'|)}{4\pi|\bm{r}-\bm{r}'|} {\bf J}(\bm{r}')d^3\bm{r}'.
	\label{eq:volume_A}
\end{equation}
Combining Eqs.~(\ref{eq:potentials}), (\ref{eq:volume_A}), the gauge condition, and comparing with Eqs.~(\ref{eq:E_volume}) one can relate tensor Green's functions with the scalar one:
\begin{equation}
	\begin{split}
	G^e_{\alpha\beta}(\bm{r}-\bm{r}') &= \left( \delta_{\alpha\beta} + \frac{1}{k_b^2}\partial_{\alpha}\partial_{\beta}  \right) g_0(\bm{r}-\bm{r}') \\
	G^m_{\alpha\beta}(\bm{r}-\bm{r}') &= \frac{1}{k_b} \xi_{\alpha\gamma\beta}\partial_{\gamma} g_0(\bm{r}-\bm{r}')
	\end{split}
	\label{eq:tensor_scalar_GF}
\end{equation}
Here differentiation is performed relative to vector $\bm{r}$.

Two-dimensional Fourier transform of the scalar Green's functions writes explicitly (see, e.g., $3.876.1,2$ and $6.667.3,4$ in \cite{Gradshteyn}, or Appendix in \cite{Munk1979}):
\begin{equation*}
	\iint\limits_{-\infty}^{\infty} d^2\bm{\rho'}\exp(-i\bm{\kappa\rho'}) \frac{\exp(ik_b|\bm{r}-\bm{r}'|)}{|\bm{r}-\bm{r}'|} = \frac{2\pi i}{k_3} \exp(-i\bm{\kappa\rho}) \exp\left( ik_3|x_3-x_3'| \right),
\end{equation*}
where $\bm{\kappa} = (k_1,\;k_2)^T$, $\bm{\rho} = (x_1,\;x_2)^T$, and $k_3^2 = k_b^2 - \bm{\kappa}^2$, $\Re k_3+\Im k_3 > 0$. Substitution of the latter equation into Eq.~(\ref{eq:volume_A}), and making the differentiation in Eq.~(\ref{eq:tensor_scalar_GF}) yield  Eqs.~(\ref{eq:Green_e}) and (\ref{eq:Green_m}). To arrive at final relations one also needs the decomposition of the $3\times3$ unit matrix
\begin{equation*}
	I=\frac{1}{k_b^2} \bm{k}^{\pm}(\bm{k}^{\pm})^T + \hat{\bm{e}}^{e\pm}(\hat{\bm{e}}^{e\pm})^T + \hat{\bm{e}}^{h\pm}(\hat{\bm{e}}^{h\pm})^T
\end{equation*}
together with transformation relations $\bm{k}^{\pm}\times\hat{\bm{e}}^{e\pm} = -k_b\hat{\bm{e}}^{h\pm}$, $\bm{k}^{\pm}\times\hat{\bm{e}}^{h\pm} = k_b\hat{\bm{e}}^{e\pm}$.


\bibliographystyle{ieeetr}
\bibliography{grating-S-matrix}

\end{document}